\documentclass[aps,pra,showpacs,twocolumn,superscriptaddress]{revtex4}
\usepackage{graphicx,color}
\usepackage{amsmath,amsthm,amsfonts,amssymb,bm}
\usepackage{times}
\usepackage{epsf}
\usepackage[colorlinks={true}]{hyperref}

\usepackage[T1]{fontenc}
\usepackage[utf8]{inputenc}
\hypersetup{citecolor={blue}, filecolor={blue}, linkcolor={blue}, urlcolor={blue}}

\newcommand{\commentold}[1]{}
\DeclareMathSymbol{:}{\mathpunct}{operators}{"3A}

\begin{document}
\title{A Measure of Non-Markovianity for Unital Quantum Dynamical Maps}
\author{S. Haseli}
\author{S. Salimi}
\email{shsalimi@uok.ac.ir}
\author{A.S. Khorashad}
\affiliation{Department of Physics, University of Kurdistan, P.O.Box 66177-15175 , Sanandaj, Iran}

\date{\today}

\begin{abstract}
One of the most important topics in the study of the dynamics of open quantum systems is the information exchange between system and environment. Based on the features of back-flow information from an environment to a system, an approach is provided to detect non-Markovianity for unital dynamical maps. The method takes advantage of non-contraction property of the von Neumann entropy under completely positive and trace preserving unital maps. Accordingly, for the dynamics of a single qubit as an open quantum system, the sign of the time-derivative of the density matrix eigenvalues of the system determines the non-Markovianity of unital quantum dynamical maps. The main characteristics of the measure is to make the corresponding calculations and optimization procedure simpler.
\end{abstract}
\pacs{03.65.Yz, 03.65.Ta, 42.50.Lc}
\maketitle
\section{Introduction}
Study of the dynamics of open quantum systems is one of the most interesting issues in the field of quantum information and computation. According to the direction of information flow (memory effects of environment), dynamics of open quantum systems  can be classified into two categories: (i) Markovian: Time evolution of a system state is determined solely by the instantaneous state of the system. In the dynamics, there is flow of information only from the system to the environment, in other words, the system smoothly loses information   (ii) Non-Markovian: The history of the system plays an important role in the time evolution which occurs along with the back flow of information from the environment to the system \cite{1}. Markovian dynamics of open quantum systems can be described by a quantum dynamical semigroup with the generator in the Lindblad form \cite{1,2}, which leads us to a time-local master equation. Since in most cases, dynamics of open quantum systems is non-Markovian, studies of non-Markovian dynamics and its effects on correlations have attracted much attention over the last few years \cite{3,4, Rajagopal, Hou, Chru,Xu,Chin,Vasile,Liu,He,Haikka1,Haikka2,Huelga,Laine,Zeng,llw,5}. Detecting non-Markovianity has been one of the interesting and challenging subjects of the study on  dynamics of open quantum systems in the last decade. There have been many attempts to introduce a general criterion for detecting non-Markovian feature of a quantum evolution. For instance, Breuer et al. introduced a significant measure based on the contraction property  of the trace distance under completely positive and trace-preserving (CPTP) maps, which is  known as BLP measure \cite{6,7}. In BLP approach, if a process is Markovian, the trace distance will be a monotonic decreasing function of time and information continuously flows from the system to the environment. In the case that the trace distance does not satisfy the above-mentioned condition, the process is obviously non-Markovian and information flows back from the environment to the system. Using the measure, one needs to do optimization over all pairs of initial states which requires complicated numerical calculations \cite{6,7}. Rivas et al. proposed a measure based on the Choi-Jamiolkowski isomorphism and divisibility  of quantum dynamical maps, which is called RHP measure  \cite{8}. In RHP method, one considers any deviation from divisibility of the corresponding dynamical map as a witness for non-Markovianity\cite{8}. The other authors use semigroup property  \cite{9}, fidelity \cite{10}, quantum fisher information \cite{11}, quantum mutual information \cite{12}, accessible information \cite{Fanchini1} and quantum loss \cite{Fanchini2} to detect non-Markovianity and to measure its degree.  In this paper, back flow of information from environment to system is regarded as a key concept to detect non-Markovianity in the case of unital quantum dynamical maps . To provide the measure,  non-contraction property of the von Neumann entropy under completely positive and trace-preserving unital (CPTPU) maps is used. Also, for the dynamics of a single-qubit as an open quantum system,  It can also be shown that the degree of non-Markovianity may be written in terms of the eigenvalues of the corresponding density matrix. In fact, in this case the non-Markovianity measure is related to non-monotonicity behaviour of the dynamics of the density matrix eigenvalues. The advantage of the measure is to make the calculations and optimization procedure simpler. It greatly increases the practical relevance of the proposed measure.
The paper is organized as follows. In Sec.II the notion of unital quantum maps is reviewed and the conditions under which a map is unital are described. In Sec.III the measure is  introduced based on the eigenvalues dynamics by using the non-contraction property of the von Neumann entropy under a CPTPU map. In Sec.IV some examples are provided to see how the measure can be applied. The conclusion is presented in Sec.V.
\section{UNITAL QUANTUM MAPS}
Let us consider an open quantum system $\mathcal{S}$ with Hilbert space $\mathcal{H}_\mathcal{S}$ and arbitrary density matrix $\rho^{\mathcal{S}}$  which  belongs to the set of  all bounded linear operators  ($B(\mathcal{H}_\mathcal{S})$) acting on the Hilbert space.
Also, let us define a CPTP map, $\Phi$, on $B(\mathcal{H}_\mathcal{S})$ which can be represented in the Kraus form\cite{1,2,13},
\begin{equation}\label{1}
  \Phi(\rho^{S})=\sum_{k}E_{k}^{\dag} \rho^{\mathcal{S}} E_{k},
\end{equation}
where $E_{k}\in B(\mathcal{H}_{s})$ and due to the trace-preserving property of the map,  $\sum_{k}E_{k}E_{k}^{\dag}=I$ is held.

{\bf \emph{Definition:}} The completely positive map $\Phi$ is unital if and only if  $\sum_{k}E_{k}^{\dag}E_{k}=I$, i.e. $\Phi$ maps the identity operator to itself  in the same space, $\Phi(I) = I$ \cite{13,14,15}. In the case of single-qubit systems, the unital maps can be expressed in terms of convex combination of the Pauli operators which are also well known as the Pauli maps \cite{14}. From geometrical point of view, the unital maps take the center of the Bloch sphere to itself. In other words, under these maps  maximally mixed states remain conserved. In the following, the unital maps are parameterized for the single-qubit case.

Let $\rho^{\mathcal{S}}$ be an operator acting on the two-dimensional Hilbert space $\mathcal{H}_\mathcal{S}=\mathbb{C}^2$ which in terms of the identity operator $I$ and the Pauli operators $\{\sigma_{x},\sigma_{y},\sigma_{z}\}$ can be written as
\begin{equation}\label{2}
 \rho^{\mathcal{S}}=\frac{1}{2}(I+\textbf{r}.{\bf{\sigma}}),
\end{equation}
where $\textbf{r}\in R^{3}$ is a Bloch vector. Every quantum  dynamical map $\Phi:B(\mathcal{H}_\mathcal{S})\rightarrow B(\mathcal{H}_\mathcal{S})$ with respect to the basis may be represented as a $4\times4$ matrix
\begin{equation}\label{3}
 L_{\Phi}=\begin{pmatrix}
1 & 0 \\
\textbf{t} &M
\end{pmatrix},
\end{equation}
in which $\textbf{t}$ is a vector in $R^3$ and $M$ is a $3\times 3$ matrix. Therefore, one can write $\Phi[\rho]=\frac{1}{2}[I+{\textbf{r}}^{\prime}.\sigma]$, where
\begin{equation}\label{4}
{\textbf{r}}^{\prime}=\textbf{t}+M \textbf{r}
\end{equation}
is an affine transformation of the Bloch vector $\textbf{r}$. In this parametrization, one can easily see that a dynamical map $\Phi$ is unital if and only if $\textbf{t}=0$ holds \cite{15}.
\section{MEASURE FOR NON-MARKOVIANITY}\label{sec}
Let us first consider a dynamical quantum process which is described by a time-local master equation of the Lindblad form
\begin{equation}\label{5}
  \dot{\rho}^{\mathcal{S}}(t)=\mathcal{L} \rho^{\mathcal{S}}(t),
\end{equation}
where $\mathcal{L}$ is a Lindblad super-operator given by \cite{1}
\begin{equation}\label{61}
\begin{split}
  \mathcal{L} \rho^{\mathcal{S}}(t)= & -i[H,\rho^{\mathcal{S}}(t)]+ \\
    & +\sum_{k} \gamma_{k}[F_{k}\rho^{\mathcal{S}}(t) F_{k}^{\dag}-\frac{1}{2}\{F_{k}^{\dag}F_{k},\rho^{\mathcal{S}}(t)\}],
\end{split}
  \end{equation}
in which $H$ is the effective  Hamiltonian, $\gamma_{k}$'s are the relaxation rates, and $F_{k}$'s are the Lindblad operators describing the type of the noise affecting the system. As long as $F_{k}$'s and $\gamma_{k}$'s are time independent, and $\gamma_{k}$'s are positive, Eq.(\ref{5}) leads to a dynamical semigroup of CPTP maps  $\Phi{(t,0)}=e^{[\mathcal{L} t]}$ with the following properties \cite{2},
\begin{description}
  \item[i)] $\Phi(t,0)$ is a dynamical map,
  \item[ii)] $\Phi(t_{1},0)\Phi(t_{2},0)=\Phi(t_{1}+t_{2},0) \;\forall\:t_{1},t_{2}\geqslant 0$,
  \item[iii)] For every $A \in B(\mathcal{H}_{\mathcal{S}})$, $Tr[(\Phi_{t} \rho^{\mathcal{S}})A]$ is a continuous function of $t$.
\end{description}

Such a dynamical quantum process characterizes a popular Markovian one. Whenever the effective Hamiltonian $H$, the Lindblad operators ($F_{k}$), and relaxation rates ($\gamma_{k}$) explicitly  depend on time,  Eq.(\ref{5}) leads to a time-dependent Markovian process, provided that all of the relaxation rates  are positive. In this case we have
 \begin{equation}\label{6}
 \begin{split}
  \mathcal{L}_{t} \rho^{\mathcal{S}}(t)= & -i[H(t),\rho^{\mathcal{S}}(t)]+ \\
     & +\sum_{k} \gamma_{k}(t)[F_{k}(t)\rho^{\mathcal{S}}(t) F_{k}^{\dag}(t)-\frac{1}{2}\{F_{k}^{\dag}(t)F_{k}(t),\rho^{\mathcal{S}}(t)\}].
 \end{split}
 \end{equation}
Here, the dynamical maps can be written in terms of a time-ordered exponential as $ \Phi(t,0)=T \exp[\int_{0}^{t} \mathcal{L}(s)ds]$. Such Markovian maps satisfy  divisibility condition \cite{8}. Divisibility condition indicates that a CPTP map ,$\Phi(t_{2},0)$, can be written as a composition of two other CPTP maps as
 \begin{equation}\label{div}
\Phi(t_{2},0)=\Phi(t_{2},t_{1})\Phi(t_{1},0).
  \end{equation}
  In some cases, it is possible that during the dynamics of the system, time-dependent relaxation rates ($\gamma_{k}(t)$) become negative in some time intervals $[t_{1},t_{2}]$. In such situations there is an intermediate dynamical map $\Phi(t_{2},t_{1})=T \exp[\int_{t_{1}}^{t_{2}} \mathcal{L}(s)ds]$ which is not CPTP. Existence of such non-CPTP intermediate dynamical map leads to the violation of the divisibility property  given by Eq.(\ref{div}). Therefore, the dynamical maps are non-Markovian\cite{8}. The criterion for non-Markovian dynamics which we are going to discuss in the following is constructed based on the monotonically increasing property of the von Neumann entropy under CPTPU maps. For this purpose, let us have a brief look at the von Neumann and quantum relative entropy concepts.
The von Neumann entropy of a quantum state with density operator $\rho^{\mathcal{S}}$  is defined as
\begin{equation}\label{f}
  S(\rho^{\mathcal{S}})=-Tr(\rho^{\mathcal{S}} \log_{2} \rho^{\mathcal{S}})=-\sum_{i} \lambda_{i} \log_{2}\lambda_{i},
\end{equation}
where $\lambda_{i}$'s are eigenvalues of   $\rho^{\mathcal{S}}$ and have a probabilistic interpretation. It indicates the lack of the knowledge about a quantum system. Since the von Neumann entropy of a pure state is zero, such states give the full knowledge about a quantum system. On the other hand, maximal mixed states with density operator $I/d$ in a $d$-dimensional Hilbert space, represent maximal ambiguity with the von Neumann entropy being equal to $\log_{2} d$.  Some of the important properties of the von Neumann entropy are:
\begin{description}
  \item[i)] The von Neumann entropy of a pure state is always minimum, $S(\rho_{pure})=0$,
  \item[ii)] For density operator $\rho$ with rank $d$ it satisfies $0\leq S(\rho)\leq \log_{2}d$,
  \item[iii)] The von Neumann entropy for an isolated quantum system does not change during the evolution, $S(\rho)=S(U \rho U^{\dag})$, and is always independent of time $d S(\rho)/dt=0$ \cite{13,15}.
\end{description}
The von Neumann relative entropy is nearly related to the von Neumann entropy and for two density operators $\rho$ and $\sigma$,  is defined as
\begin{equation}\label{rela}
S(\rho \Vert \sigma)=tr(\rho \log_2 \rho)-tr(\rho \log_2 \sigma).
\end{equation}
The von Neumann relative entropy introduced in above is not a distance in the mathematical sense, since it does not satisfy the triangle inequality and it is not symmetric \cite{13}.
The von Neumann relative entropy is contractive and non-increasing under CPTP maps, i.e. if $\Phi$ is a CPTP map , we will have
\begin{equation}\label{8}
S \left(\Phi(\rho^{\mathcal{S}})\parallel \Phi(\sigma^{\mathcal{S}}) \right)\leq S\left(\rho^{\mathcal{S}}\parallel \sigma^{\mathcal{S}}\right).
\end{equation}
 Considering this property of the von Neumann relative entropy, one can easily find that the von Neumann entropy is non-contractive under CPTPU maps. In order to prove this statement, suppose that $\sigma^{S}=\frac{1}{d} I$, thus Eq.(\ref{rela}) can be written as
\begin{equation}\label{g}
  S(\rho^{\mathcal{S}}\parallel\frac{1}{d}I)=-S(\rho^{\mathcal{S}})+\log_2 d.
\end{equation}
Substituting  Eq.(\ref{g}) into Eq.(\ref{8}) and after some straightforward calculations, one can easily find that the von Neumann entropy is non-contractive under CPTPU maps  \cite{15, non},
\begin{equation}\label{10}
S\left(\Phi(\rho^{\mathcal{S}})\right) \geq S\left(\rho^{\mathcal{S}}\right).
\end{equation}
It indicates that the purity of the system decreases under the above-mentioned dynamical maps.
Regarding this interpretation, one can introduce a witness for non-Markovianity:
a unital quantum dynamical map $\Phi{(t,0)}:\rho^{\mathcal{S}}(0)\rightarrow \rho^{\mathcal{S}}(t)=\Phi{(t,0)} \rho^{\mathcal{S}}(0)$ is non-Markovian if
\begin{equation}\label{12}
 \frac{d}{dt}S\left(\rho^{\mathcal{S}}(t)\right)< 0.
\end{equation}
The non-monotonicity property of the von Neumann entropy of an open quantum system can be interpreted as back flow of information from environment to the system, so any deviation from the monotonically increasing property of the von Neumann entropy shows that the dynamical maps are not completely positive and consequently are non-Markovian. According to this criterion for the non-Markovian dynamics of open quantum systems, one can introduce a new method to quantify the degree of non-Markovianity. Mathematically, the measure can be written as
\begin{equation}\label{degree}
N_{S}=\max_{\lbrace\rho^{\mathcal{S}}(0)\rbrace}\int_{\frac{d}{dt}S\left(\rho^{\mathcal{S}}(t)\right)< 0} dt \;\frac{d}{dt}S\left(\rho^{\mathcal{S}}(t)\right).
\end{equation}
The time integration is taken over all time intervals $(t_{i},t_{j})$ on which the time derivative of the von Neumann entropy is negative, and the maximization is evaluated over all possible initial states $\rho^{\mathcal{S}}(0)$ of the system. The composition law of divisibility given in Eq.(\ref{div}) implies that this measure vanishes for all divisible unital quantum dynamical maps, that is, according to this measure all divisible unital dynamical maps define Markovian processes. In order to prove this statement, assume that the CPTPU map $ \Phi(t,0)$is divisible meaning that for all $t, \tau\geq 0 $ one has
\begin{equation}\label{div2}
\Phi(t+\tau,0)=\Phi(t+\tau,t)\Phi(t,0),
\end{equation}
where  $\Phi(t+\tau,t)$ is also a CPTPU map. Therefore, for any initial state $\rho^{\mathcal{S}}(0)$ we have
\begin{equation}\label{monoto}
\rho^{\mathcal{S}}(t+\tau)=\Phi(t+\tau,t)\rho^{\mathcal{S}}(t).
\end{equation}
Due to the fact that $\Phi(t+\tau,t)$ is a CPTPU map, one can show the non-contraction property, Eq.(\ref{10}), as
\begin{equation}\label{mon}
S(\rho^{\mathcal{S}}(t+\tau))\geq S(\rho^{\mathcal{S}}(t)).
\end{equation}
Thus, for all divisible unital quantum dynamical maps the von-Neumann entropy monotonically increases and $N_{S}=0$. In other words, we can say that all divisible unital quantum dynamical maps are Markovian. However, the inverse statement is not necessarily true, that is, there are non-divisible unital quantum dynamical maps for which the entropy does not show any temporary decrease at all. In the case of unital maps
according to the above-mentioned points non-Markovianity must be described by non-divisible unital dynamical maps. In Markovian processes, the state of an open quantum system loses its purity due to interaction of the system with the surrounding environment which gives rise to increase the value of the von Neumann entropy. The increase of the von Neumann entropy during the dynamics of the system can be interpreted as the loss of the system information. Conversely, increasing the purity of the state of an open quantum system is equivalent to the reduction of the von Neumann entropy. This reduction can be considered as back flow of information from the environment to the system which is the main characteristics of the non-Markovian dynamical quantum processes. We will continue our discussion by looking at the dynamics of a one-qubit system due to its interaction with its surrounding environment. For one-qubit model the density matrix at time $t$ can be obtained by
\begin{equation}
  \rho^{\mathcal{S}}(t)=\Phi(t,0) \rho^{\mathcal{\mathcal{S}}}(0)=\begin{pmatrix}
\rho^{\mathcal{S}}_{11}(t) & \rho^{\mathcal{S}}_{12}(t) \\
\rho^{\mathcal{S}}_{21}(t) &\rho^{\mathcal{S}}_{22}(t)
\end{pmatrix}.
\end{equation}
The corresponding von Neumann entropy can be calculated as $S(\rho^{\mathcal{S}}(t))=-\lambda_{+}(t)\log_{2}\lambda_{+}(t)-\lambda_{-}(t)\log_{2}\lambda_{-}(t)$, in which  $\lambda_{\pm}(t)$ are the eigenvalues of $\rho^{\mathcal{S}}(t)$ obtained as
\begin{equation}
  \lambda_{\pm}(t)=\frac{1\pm\sqrt{1-4(\rho^{\mathcal{S}}_{11}(t)\rho^{\mathcal{S}}_{22}(t)-|\rho^{\mathcal{S}}_{12}(t)|^{2})}}{2}.
\end{equation}
By taking the time derivative of the von Neumann entropy,
\begin{equation}\label{13}
  \frac{d}{dt}S(\rho^{\mathcal{S}}(t))=\frac{d\lambda_{\pm}(t)}{dt}\log_{2}\frac{\lambda_{\mp}(t)}{\lambda_{\pm}(t)},
\end{equation}
it can be shown from Eq.(\ref{12})  that the dynamics of the system is non-Markovian, if $\eta_{-}=d\lambda_{-}(t)/dt<0$ or  $\eta_{+}=d\lambda_{+}(t)/dt>0$. Regarding the relation $\lambda_{+}(t)=1-\lambda_{-}(t)$ and Eq.(\ref{degree}),  the degree of non-Markovianity of the dynamical maps for one-qubit models can be defined as
\begin{equation}\label{14}
\begin{split}
  N_{e}(\Phi)= & \max_{\{\rho^{\mathcal{S}}(0)\}}\int_{\eta_{+}>0}\eta_{+} dt \\
    &= -\max_{\{\rho^{\mathcal{S}}(0)\}}\int_{\eta_{-}<0}\eta_{-} dt.
\end{split}
\end{equation}
The integral is taken over all time intervals $t\in(a_{i},b_{i})$ on which for the first equality in Eq.(\ref{14}) $\eta_{+}$ is positive, and for the latter $\eta_{-}$ is negative. The maximization is evaluated over the all input states of the one-qubit system  $\{\rho^{\mathcal{S}}(0)\}$.  Like the  measures proposed by the other authors, this measure depends on the initial state. The general problem one faces with in applying the existing non-Markovian measures is to perform the optimization procedure. Fortunately, from Eq.(\ref{14}) one observes that the optimization can be taken over the all initial states of the one-qubit system, which  simplifies calculations. In view of $\eta_{+}=d\lambda_{+}(t)/dt$ and $\eta_{-}=d\lambda_{-}(t)/dt$, the non-Markovianity measure in Eq.(\ref{14}) can be
rewritten as
\begin{equation}\label{15}
\begin{split}
 N_{e}(\Phi)= & \max_{\{\rho^{\mathcal{S}}(0)\}}\sum_{i} (\lambda_{+}(b_{i})-\lambda_{+}(a_{i})) \\
    &=-\max_{\{\rho^{\mathcal{S}}(0)\}}\sum_{i} (\lambda_{-}(b_{i})-\lambda_{-}(a_{i})).
\end{split}
\end{equation}
To calculate the degree of non-Markovianity via the first equality in Eq.(\ref{15}), for any initial state one has to specify  the total increase of the greater eigenvalue $\lambda_{+}(t)$ over each time interval $(a_{i},b_{i})$, on which  $\lambda_{+}(t)$ is an ascending function, and sum up the contributions of all such intervals. Finally, one can find  $ N_{e}(\Phi)$ by determining the maximum over the all initial states $\{\rho^{\mathcal{S}}(0)\}$.
\section{SOME EXAMPLES IN DYNAMICAL MODELS}
\subsection{Phase Damping Dynamical Model}
\begin{figure}[h]
  \centering
  \includegraphics[width=7cm]{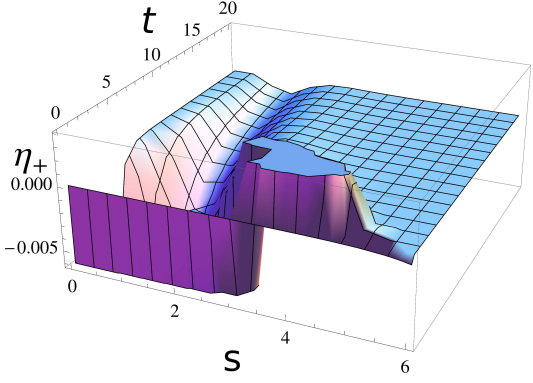}\\
  \caption{Behaviour of $\eta_{+}$ as a function of time  $t$ and $s$.}\label{fig1}
\end{figure}

Let us consider a two-level system which interacts with a bosonic environment, where the type of the interaction is the pure dephasing. In this case, the dynamics of one-qubit system is captured by the following time-local master equation \cite{12,16}
\begin{equation}\label{18}
  \mathcal{L}_{t}(\rho^{\mathcal{S}}(t))=\frac{\gamma(t)}{2}\left(\sigma_{z}\rho^{\mathcal{S}}(t)\sigma_{z}-\rho^{\mathcal{S}}(t)\right),
\end{equation}
where $\sigma_{z}$ is the Pauli spin operator in the $z$-direction, and $\gamma(t)$ is the time-dependent dephasing rate. The result of this type of interaction between one-qubit system and the bosonic environment is the decay of the off-diagonal elements with the decoherence factor $e ^{\Gamma(t)}$, where  $\Gamma(t)\geq 0$ always holds. When temperature of the environment is zero, $\Gamma(t)$ can be defined by
 \begin{equation}\label{20}
  \Gamma(t)=4\int d\omega J(\omega) \frac{1-\cos\omega t}{\omega^{2}},
\end{equation}
where $J(\omega)$ is the spectral density of the environment \cite{16}. We assume that the
spectral density of the environment is Ohmic-like \cite{17},
\begin{equation}\label{23}
  J(\omega)=\omega_{c}^{1-s}\omega^{s}e^{-\frac{\omega}{\omega_{c}}},
\end{equation}
where $\omega_{c}$ and $s$ are the cutoff frequency and Ohmicity parameter, respectively. Regarding the different values of $s$, one can have sub-Ohmic ($s<1$),  Ohmic ($s=1$) and super-Ohmic ($s>1$) spectral densities. Further details about this model are presented in the appendix. The effect of the phase damping dynamical quantum map on the initial density matrix of the one-qubit system can be written as
\begin{equation}\label{19}
  \Phi{(t,0)}\rho^{\mathcal{S}}(0)=\begin{pmatrix}
\rho^{\mathcal{S}}_{11}(0) & \rho^{\mathcal{S}}_{12}(0)e^{-\Gamma(t)} \\
\rho^{\mathcal{S}}_{21}(0)e^{-\Gamma(t)} &\rho^{\mathcal{S}}_{22}(0)
\end{pmatrix}.
\end{equation}
Time derivative of the eigenvalues of the above density matrix is straightforwardly obtained as
\begin{equation}\label{21}
\begin{split}
 \frac{d\lambda_{+}(t)}{dt}= & -\frac{d\lambda_{-}(t)}{dt}= \\ &=\frac{-8\gamma(t)e^{-2\Gamma(t)}|\rho^{\mathcal{S}}_{12}(0)|^{2}}{4\sqrt{1-4(\rho^{\mathcal{S}}_{11}(0)\rho^{\mathcal{S}}_{22}(0)-e^{-2\Gamma(t)}|\rho^{\mathcal{S}}_{12}(0)|^{2})}},
\end{split}
\end{equation}
where the first equality is the result of trace preserving of the dynamical map. Regarding the non-Markovianity condition which has been presented instantly after Eq.(\ref{13}), i.e. $\eta_{+}>0$ and $\eta_{-}<0$, one can find that the phase damping dynamical map is non-Markovian if $\gamma(t)<0$, which agrees with the result obtained by RHP measure in \cite{6,8}.
 The maximum in Eq.(\ref{14}) is achieved by the initial states $\rho^{\mathcal{S}}(0)=|\pm\rangle\langle\pm|$, where $|\pm\rangle=\frac{1}{\sqrt{2}}(|0\rangle \pm|1\rangle)$. Fig.(\ref{fig1}) shows the behaviour of the time derivative of the greater eigenvalue,  $\eta_{+}$, for these  states as a function of time $t$ and parameter $s$. As can be seen, for some intervals, $s\in[2.5,5.5]$, the process is non-Markovian due to the positivity of $\eta_{+}$. In other words, the non-Markovian behaviour appears when the qubit interacts with a super-Ohmic reservoir. Positivity of $\eta_{+}$ (negativity of $\eta_{-}$) means that the probability of finding the state of the system in the initial state increases in some time intervals during the process, which is interpreted as the back flow of information from the environment to the system.
 The degree of the non-Markovianity of the pure dephasing dynamical map, $N_{e}$, is plotted as a function of Ohmicity parameter $s$ in Fig.(2). The results agree with those obtained by BLP measure \cite{addis}.
\begin{figure}[h]
  \centering
  \includegraphics[width=7cm]{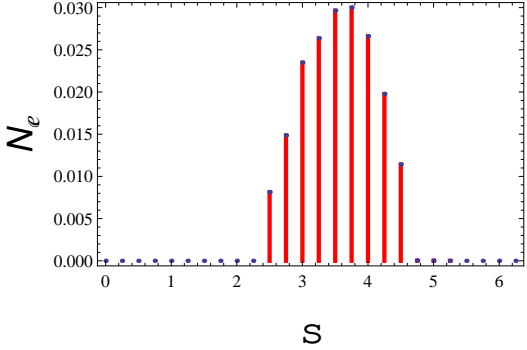}
 \center{ \caption{Degree of non-Markovianity for different types of reservoir, sub-Ohmic ($s<1$), Ohmic ($s=1$), super-Ohmic ($s>1$).}}\label{s}
\end{figure}
\subsection{Dephasing Model With Colored Noise}
Let us consider a two-level quantum system which interacts with an environment having the properties of random telegraph signal noise. The model was first presented by Daffer et al.\cite{19}. In this model, a dynamical quantum map is described by a master equation of the form
\begin{equation}\label{master}
 \dot{\rho}^{\mathcal{S}}(t)=K \mathcal{L} \rho^{\mathcal{S}}(t),
 \end{equation}
 where $K$ is a time-dependent integral operator defined as $K \psi = \int_{0}^{t}k(t-\acute{t})\psi(\acute{t})d \acute{t}$, where $k(t-\acute{t})\psi(\acute{t})$ is the kernel function which determines the type of memory in the environment. In order to study the master equation in the form of Eq.(\ref{master}), one can consider a time-dependent Hamiltonian as
\begin{equation}
  H(t)=\sum_{k} \Gamma_{k}(t) \sigma_{k},
\end{equation}
where, $\sigma_{k}$ is the Pauli spin operator in the $k$-direction and $\Gamma_{k}(t)$'s are  random variables obeying the statistics of a random telegraph signal. $\Gamma_{k}(t)$ is related to the variable $n_{k}(t)$ by  $\Gamma_{k}(t)=a_{k} n_{k}(t)$, in which $n_{k}(t)$ has a Poisson distribution with a mean being equal to $t/2\tau_{k}$, and $a_{k}$'s are coin-flip random variables possessing the values $\pm a_{k}$. Making use of the von Neumann equation, $\dot{\rho}^{\mathcal{S}}(t)=-(i/\hbar)[H(t),\rho^{\mathcal{S}}]$, one can find an expression for the density matrix of the two-level system as
 \begin{equation}\label{z}
  \rho^{\mathcal{S}}(t)=\rho^{\mathcal{S}}(0)-i\int_{0}^{t}\sum_{k} \Gamma_{k}(t)[\sigma_{k},\rho^{\mathcal{S}}(s)]ds.
 \end{equation}
Substituting Eq.(\ref{z}) back into the von Neumann equation and performing the stochastic averages lead to the following master equation
 \begin{equation}\label{z1}
   \dot{\rho}^{\mathcal{S}}(t)=-\int_{0}^{t} \sum_{k} e^{-(t-\acute{t})/\tau_{k}}a_{k}^{2}[\sigma_{k},[\sigma_{k},\rho^{\mathcal{S}}(\acute{t})]]d\acute{t},
 \end{equation}
 where the kernel function comes from the correlation functions of random variables $\langle \Gamma_{j}(t) \Gamma_{k}(\acute{t})\rangle=a_{k}^{2}\exp(-|t-\acute{t}|/\tau_{k})\delta_{j,k}$.
 As one can see, Eq.(\ref{z1}) is in the form of Eq.(\ref{master}). Also, Daffer et al. showed that the dynamical process described by Eq.(\ref{z1}) is completely positive when two of the $a_{k}$ are zero, i.e. the random telegraph noise only acts in one direction. Whenever, $a_{1}=a_{2}=0$ and $a_{3}=a$ the dynamical process is known as a completely positive  dephasing  with the colored noise. Thus one can show this quantum process of the two-level system by a CPTP map in the Kraus form as follows \cite{13}
\begin{equation}\label{kraus}
   \rho^{\mathcal{S}}(t)=\sum_{i}A_{i}\rho^{\mathcal{S}}(0)A_{i}^{\dag},
 \end{equation}
 where $A_{i}$'s are the Kraus operators describing the dynamics of the system and are given by

 \begin{equation}\label{ko}
   A_{1}=\sqrt{\frac{1+\Lambda(\nu)}{2}}\:I, \quad
   A_{2}=\sqrt{\frac{1-\Lambda(\nu)}{2}}\:\sigma_{z},
 \end{equation}

 where $\Lambda(\nu)=e^{-\nu}(\cos(\mu\nu)+\sin(\mu\nu)/\mu)$, $\mu=\sqrt{(4a\tau)^{2}-1}$, and $\nu=t/2\tau$ is dimensionless time.

Next by straightforward calculations the time derivative of the greater eigenvalues of the density matrix $\rho^{\mathcal{S}}(t)$ in Eq.(\ref{kraus}),  $\eta_{+}$, can be obtained as
 \begin{equation}\label{ez}
   \frac{d\lambda_{+}(t)}{dt}=\frac{d\Lambda(\nu)}{d\nu}\frac{2|\rho^{\mathcal{S}}_{12}(0)|^{2}}{\sqrt{1-4(\rho^{S}_{11}(0)\rho^{\mathcal{S}}_{22}(0)-\Lambda(\nu)^{2}|\rho^{\mathcal{S}}_{12}(0)|^{2})}\tau}
 \end{equation}

\begin{figure}[h]
  \centering
  \includegraphics[width=7cm]{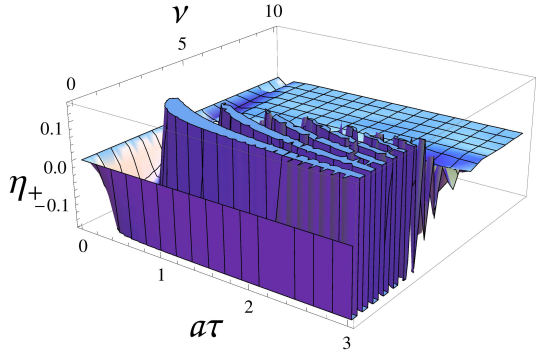}\\
  \caption{Variation of $\frac{d\lambda_{+}(t)}{dt}$ as a function of time and $a\tau$.}\label{fig3}
\end{figure}

\begin{figure}[h]
  \centering
  \includegraphics[width=7cm]{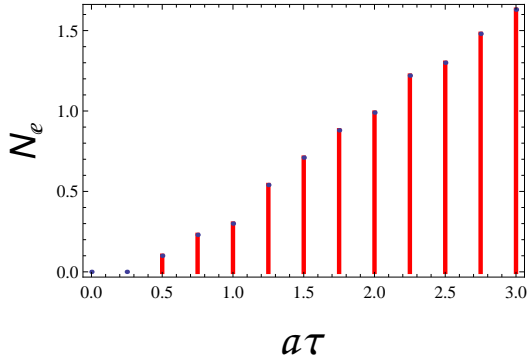}\\
  \caption{ Degree of non-Markovianity in terms of the fluctuation rate, $a\tau$.}\label{fig4}
\end{figure}

Considering the non-Markovianity condition introduced in Sec.\ref{sec}, the dephasing dynamics of the system with the colored noise is non-Markovian if $d\Lambda(\nu)/d\nu>0$,  which agrees with the result obtained by applying some recently introduced measures \cite{6,12}. In Fig.(\ref{fig3}), the time derivative of the greater eigenvalue, $\eta_{+}(t)$, is plotted as a function of time and $a\tau$ for the initial states $\rho^{\mathcal{S}}(0)=|\pm\rangle\langle\pm|$ ($|\pm\rangle=\frac{1}{\sqrt{2}}[|0\rangle \pm|1\rangle]$). Here, we emphasize that these initial states come from maximization procedure in Eq.(\ref{14}). As can be seen, if $a\tau\geq\frac{1}{2}$, then in some time intervals $\eta_{+}(t)$ may be positive, which means that the dynamics of the system is non-Markovian. Also, the degree of non-Markovianity, $N_{e}$ in Eq.(\ref{14}), is plotted in Fig.(\ref{fig4}).

 From the above examples, it can clearly be seen that the structure of reservoir affects the non-Markovianity character of the dynamics. In phase damping dynamical model, non-Markovianity is revealed in the super-Ohmic regime and in dephasing model with colored noise, the degree of non-Markovianity increases by increasing the fluctuation rate of the external field.

\section{CONCLUSIONS}
In this paper a non-Markovianity measure, $N_{e}$, has been proposed which is expressly connected to the time derivative of the density matrix eigenvalues of an open quantum system. In the measure, the back flow of information  from the environment to the system has been used as a feature of a non-Markovian process. It has been shown that if the time derivative of the greater eigenvalue is positive (or that of the smaller one is negative), the purity of the system state increases during the time evolution with respect to that of the initial state. This means that the information flows from the environment to the system and therefore the dynamics is non-Markovian. In addition, it has also been shown that the structure of the reservoir affects on the Markovianity and non-Markovianity character of an open quantum system dynamics \cite{20}. The advantage of this measure is its simplicity in calculations and optimization procedure which is only taken over the all initial input states of the single qubit system.
 Although in this paper we just only focus on CPTPU maps, it should be pointed out that applying the measure for this kind of maps is simpler than other measures. This is due to the fact that in this measure we do not require
complicated statistical methods, such as Monte Carlo sampling of the pairs of the initial states in order to do optimization procedure.
\section*{APPENDIX}
The pure dephasing interaction between a two-level system and surrounding bosonic environment is given by
\begin{equation}\label{16}
 H=\frac{\omega_{0}}{2}\sigma_{z}+\sum_{k}\omega_{k}b_{k}^{\dag}b_{k}+\sum_{k}\sigma_{k}(g_{k}b_{k}^{\dag}+g_{k}^{\ast}b_{k}),
\end{equation}
where $\sigma_{z}$ is the usual Pauli matrix in the $z$-direction, $\omega_{0}$ is the two-level system frequency, the $b_{k}$($b_{k}^{\dag}$) are the annihilation(creation) operators which satisfy the commutation relation $[b_{k},b^{\dag}_{\acute{k}}]=\delta_{k,\acute{k}}$, $g_{k}$ is a constant which can control the strength of the coupling between the system and the environment. In this model the off-diagonal elements of the density matrix of the two-level system decay during the quantum process, while the diagonal elements are constant in time because  there is no transition between energy levels which is due to this fact that   $[H,\sigma_{z}]=0$ holds.
In interaction picture the Hamiltonian is obtained as
\begin{equation}
H_{I}(t)=\sum_{K}\sigma_{z}(g_{k}a_{k}^{\dag}e^{i\omega_{k}t}+g_{k}^{\ast}a_{k}e^{-\omega_{k}t}).
\end{equation}
When the system interacts with a large environment, one can work in the continuum limit and  have a replacement $\sum_{k} \vert g_{k} \vert^{2}\longrightarrow \int d\omega J(\omega) \delta(\omega_{k}-\omega)$, where $J(\omega)$ is the spectral density of the environment \cite{16,17}. Using the second order time-convolutionless master equation \cite{1} at zero temperature, one can find the master equation appeared in Eq.(\ref{18}) . This model can be described in the Kraus representation form as
\begin{equation}\label{sor}
  \rho^{\mathcal{S}}(t)=\sum_{i=1}^{2} D_{i}(t)\rho^{\mathcal{S}}(0)D_{i}^{\dag}(t),
   \end{equation}
   where the Kraus operators $D_{i}(t)$ are given by
   \begin{equation}\label{g8}
   D_{1}(t)=\sqrt{\frac{1+e^{-\Gamma(t)}}{2}} \: I, \quad D_{2}(t)=\sqrt{\frac{1-e^{-\Gamma(t)}}{2}}\: \sigma_{z}.
 \end{equation}



\end{document}